\documentstyle[11pt,aaspp4]{article}

\begin{document}
 
\title{Canada-France Redshift Survey XI: Morphology of high-redshift 
field galaxies from high-resolution ground-based imaging}
 
\author{David Schade\footnotemark[1] \& S. J. Lilly\footnotemark[1] }
\affil{Department of Astronomy, University of Toronto, Toronto, Canada}

\and

\author{O. Le F\`evre\footnotemark[1] \& F. Hammer\footnotemark[1] }
\affil{DAEC, Observatoire de Paris Meudon, 92195 Meudon CEDEX France}

\and

\author{D. Crampton\footnotemark[1]}
\affil{Dominion Astrophysical Observatory, Victoria, Canada}

\footnotetext[1]{Visiting Astronomer, Canada-France-Hawaii Telescope, 
which is operated by the National Research Council of Canada, the Centre
Nationale de la Recherche Scientifique of France, and the University
of Hawaii.}
 
\begin{abstract}

 The 143 galaxies with secure redshifts 
($z_{median}=0.62$) from the  1415+52
field of the Canada-France Redshift Survey has been imaged 
with median seeing of 0.67 arcseconds
(FWHM). Structural parameters have been derived by fitting multi-component
models and the results confirm two phenomena seen in a smaller
sample of galaxies imaged with the {\it Hubble Space Telescope}. 

First, $11\pm3\%$ of the galaxies 
lie off the normal locus of colour vs. bulge fraction $B/T$.
This class
of objects  (``blue-nucleated galaxies'' or BNGs) was identified
using HST observations (Schade et al. 1995, CFRS IX) and it was shown
that they are associated with peculiar/asymmetric structure and 
merger/interactions. 
 The observed frequency of BNGs in this sample is $14 \pm 4\%$ $0.5 < z < 1.2$ 
and $6^{+6}_{-3}\%$ at $0.2 < z < 0.5$ but the true frequency is
likely to be a factor $\sim2$ higher after corrections are made for
the effect of asymmetric/peculiar structures.

 Galaxy disks at $0.5 < z < 1.1$ are found to
 have a mean rest-frame, inclination-corrected central
surface brightness of $\mu_{AB}(B)=19.8 \pm 0.1$
 mag arcsec$^{-2}$, 
$\sim 1.6$ mag  brighter 
than the Freeman (1970) value. 
At low redshift ($0.2 < z < 0.5$) the mean surface brightness
($\mu_{AB}(B)=21.3 \pm 0.25$) is consistent with the Freeman
value. 

 These results are consistent with the $HST$ observations. 
With larger 
numbers of galaxies and therefore more statistical weight they demonstrate
the capabilities, and limits, of ground-based work in the study of
 galaxy morphology at high redshift.

\end{abstract}


\section{INTRODUCTION}

The morphological class of a galaxy 
correlates
well with its physical properties: HI content, stellar population
via color, specific angular momentum (e.g. Roberts \& Haynes 1994), 
present and past average
star-formation rates (Kennicutt, Tamblyn, \& Congdon 1994). Principal component analysis 
 indicates that,
for both disk and elliptical galaxies, roughly 90\%
of the variance in galaxy properties is
accounted for by only two variables which are related to form
(or morphology) and scale (luminosity or size), 
respectively (Okamura, Watanabe, and Kodaira 1988). 
Fortunately, both of these quantities
can be measured by imaging alone at large
distances and, furthermore,
 these properties are closely related
to the formation process 
and thus likely to be keys to understanding the
formation and evolution of distant galaxies.
 
 Hubble (1926) devised our basic system of galaxy classification, 
 and this was extended
by deVaucouleurs (1959) and by Sandage (1961). The morphological
type of a galaxy is determined by a number of properties
including the bulge-to-disk luminosity ratio and the tightness
and degree of resolution of spiral arms.  The
classification has historically done from rest frame
blue band images. The 
process is subjective, but it is repeatable and there is a good measure
of agreement between different workers (Lahav 1995, Buta et al. 1994).

 The ability to classify galaxies by Hubble type
visually from ground-based images is severely
compromised as redshift increases, 
and the process is impossible
at the redshifts investigated here because much of the
detail is smoothed away by the effects of seeing. 
Nevertheless, quantitative information can be obtained 
from the light distribution and a
classification based on 
fractional bulge luminosity (larger in early-type galaxies)
can be done at high redshift. 

 The Canada-France Redshift Survey (CFRS) provides a sample
of $ 591$ galaxies with$17.5 \le I_{AB} \le 22.5$ and 
measured redshifts 
(see Le F\`evre et al. 1995 CFRS II,
Crampton et al. 1995 CFRS V, and references therein).
The high median redshift
($z=0.6$)  together with the $I$ band (8300 \AA
=rest frame $B$ at $z=0.9$) selection makes
it possible to construct high-redshift samples of galaxies
that are directly comparable to the local population.
 The evolution of
the tri-variate galaxy luminosity function 
$\phi (M,z,color)$ is discussed by Lilly et al. 1995 (CFRS VI). 
The luminosity
function of the red population
evolves very little whereas 
that of the blue galaxies varies 
strongly with redshift. The luminosity function is
a statistical description of the population  and
does not imply any particular physical interpretation.
 The observed behaviour of the blue population can
thus be equally well described as a uniform brightening of
individual galaxies with increasing redshift or as the presence 
of a numerous new population that is absent at the
present day. Physical information about individual galaxies,
such as morphology or kinematics, is needed to break this
degeneracy.

 The fine resolving power of the Hubble Space Telescope makes it
uniquely powerful for studying high-redshift galaxies. Work
is in progress (e.g., Dressler et al. 1994, Griffiths et al. 1994a)
but,to date, few redshifts exist for faint ($I_{AB}\sim 22$),
high-redshift ($z > 0.5$) field galaxies 
that have been observed with HST. Schade et al. 1995 (CFRS IX) present
results for a sample of 32 galaxies with $0.5 \le z \le 1.2$ from
two fields of the CFRS, the first analysis based on secure redshifts
from a carefully selected sample.

The present work relies critically on the excellent
seeing at Mauna Kea, and exploits the wide field capabilities 
of ground based imaging cameras to
establish the basic morphological properties of larger numbers 
of high-z galaxies in the CFRS.
Although the resolution is much lower than $HST$, at the present time
this is the largest sample with secure redshifts and is thus 
an important preliminary to large numbers
of observations of $z>0.5$ galaxies with HST.

 In this paper we discuss the modelling of the two-dimensional
luminosity distributions of high-redshift galaxies with
functions that have been found to be appropriate for nearby
objects (\S 2) and present simulations to demonstrate the
feasibility of such an approach using ground-based imaging
(\S 3). Our observations for a sample of 195 objects (both stars
and galaxies) are described in \S 4 along with the results
of the model-fitting. In \S 5 we discuss the relation between
the morphological results and the evolving luminosity function and
compare our work with previous results including our own $HST$
imaging. Photometry is presented in the ${AB}$ systems
($B_{AB}=B-0.17$, $(U-V)_{AB}=(U-V)+0.7$), and it is assumed
throughout that $H_\circ=50$ km sec$^{-1}$ Mpc$^{-1}$ and 
$q_{\circ}=0.5$.

\section{MODELLING HIGH-Z GALAXIES}

\subsection{Choice of models}

Our choice of models is guided by the two considerations.
 First, the fitted parameters must be
comparable to morphological parameters that have been derived from
nearby galaxies and, second, it must be possible to extract these 
parameters reliably for high-redshift
galaxies. The resolution in
physical units in the frame of the galaxies is much smaller
in this study than in similar local studies so that these
two considerations are not trivial. The best approach is
to use the simplest models that have been used to fit the luminosity profiles
of nearby galaxies.

 A discussion of techniques for photometric decomposition
of galaxies into distinct components is given by Simien (1988).
Few systematic studies have been carried out on samples
larger than 50 galaxies. 
 Surface photometry on small 
samples of galaxies has been done
 by Freeman (1970),
 Kormendy (1977b),
 Burstein (1979a,1979b) , 
 Boroson (1981), and analyses of larger samples
have been done by van der Kruit (1987), Kent (1985),
and Kodaira, Watanabe,
and Okamura (1986).
These references include
discussions of the relative merits of various fitting functions
and the difficulties of decomposing multi-component
luminosity profiles. 

Most of the above studies use 
the deVaucouleurs $r^{1/4}$ law to describe the compact 
spheroidal components
and an exponential profile to fit the disk components.
Clearly the gross structure of most galaxies can be described
by these models.  On the other hand,
all studies find some anomalous galaxies that cannot be
well-fitted by standard models. The fraction of such galaxies
ranges from about 25\% (poor or impossible fits from Kent 1985)
 to 13\% (Kodaira, Wanatabe, and Okamura
1986) and the failures are caused by the presence of structures 
such as bars, dust lanes,
spiral structure, and galaxy cores which are
not accommodated by the simple models adopted here.
 The most severe difficulty
in a study like the present one is not the statistical
uncertainties in the model-fitting process but the likelihood
of systematic errors due to these additional structures. 
Nevertheless, the majority
of the measurements will be meaningful if the high-redshift
galaxy population has a similar distribution of
morphological properties to the local population. In particular, 
 the gross size measurements of disks will be largely
unaffected by moderate peculiarities in the luminosity
profiles.
Disk-dominated galaxies constitute 60\%---80\% of the field
galaxy population (Buta et al. 1994) and the typical disk size
 ($\sim$ few kpc 
scale length or 0.35
arcseconds at $z=0.75$) is large enough to be
measured
using ground-based imaging with excellent seeing.

\subsection{ Determining the point-spread function}

 The typical scale size of the galaxies discussed here 
 is smaller
than the  FWHM of the point-spread function
so that a careful
treatment of the convolution with seeing 
is an important part of the present analysis.
  Empirical point-spread functions (PSFs) formed the basis of the 
analysis using DAOPHOT software (Stetson 1987) to construct and
manage the PSFs. The construction of the PSF followed usual
procedures for stellar photometry (Stetson 1987) using a number of
bright stars, fitting and subtracting, examining the residuals to
exclude stars with near neighbors or other flaws.  The PSF varied 
little with position on the CCD whereas frame-to-frame
variations were large. 
The radius over which the PSF was defined was typically
5 arcseconds.

\subsection{Details of the fitting procedure}

 An idealized, continuous galaxy model is transformed into the
observational space by defining the galaxy center, 
orientation, and scale with respect to the detector pixel grid.
The galaxy model is convolved with the point-spread function
and the resulting luminosity distribution is integrated 
over each pixel to produce the expected number of counts in
each pixel of the observed image, given the model parameters.

 In practice, a discrete point-spread function (PSF) is used
(it can be derived directly from the data thereby dispensing
with the requirement that it conform to some analytical
form) and the continuous galaxy model is integrated
over each detector pixel. The convolution is then executed
using discrete fast fourier transforms.
The bulge component is represented by a deVaucouleurs $R^{1/4}$
law:
\begin{equation}
I_B(r_B)=I_B(0){\rm exp} \biggl[ -7.67 \biggl({r_B\over r_e}\biggr)^{0.25} \biggr] 
\end{equation}
and the disk component by:
\begin{equation}
I_D(r_D)=I_D(0) {\rm exp} \biggl({r_D\over h }\biggr) 
\end{equation}
where I(0) is the central surface brightness, $r_e$ is the bulge
effective (or half-light) radius, and $h$ is the disk scale
length. If  $(x_c,y_c)$ give the position of the galaxy center,
then at a position $(x,y)$, we have $dx=x-x_c$ and $dx=y-y_c$,
\begin{equation}
dx_B=dx*{\rm cos}(\theta_B) +dy*{\rm sin}(\theta_B) \newline
\end{equation}
\begin{equation}
dy_B=\bigl(-dx*{\rm sin}(\theta_B) +dy*{\rm cos}(\theta_B)\bigr)/ar_B
\end{equation}
and
\begin{equation}
r_B^2=dx_B^2 + dy_B^2
\end{equation}
where $\theta_B$ is the position angle of the major axis of
the bulge component and $ar_B$ is the axial ratio (minor/major) of
the bulge. A similar equation holds for the disk component. The position
angles of the two components are allowed to vary independently.

In order to minimize sampling errors,
the PSF  is shifted  
to the position of the centroid
of the galaxy.
 A fixed-precision Simpson's rule integration scheme 
 evaluates each integral to a minimum
precision of 1 part in a thousand. 
Multiple images of an object (as many as 30 in our data) are paired with their
corresponding PSFs and all fitted simultaneously. In this way,
images of an object obtained with different instruments and
filters (including parameters accounting for the difference
in color between bulge and disk) can be efficiently analysed.
Weighting is strictly according
to Poisson statistics. Pixels affected by cosmic rays were
excluded from the fits.
 A Levenberg-Marquardt non-linear fitting 
algorithm  (Press, Teukolsky, Vetterling, \& Flannery 1992) which proved robust, was used to locate
the minimum in $\chi^2$. Errors were estimated from the covariance matrix
and verified using simulations.

\section{RESULTS OF SIMULATIONS}

Extensive simulations
were done to gain an understanding 
of both the parameter space being investigated and the properties
of the fitting procedure. 
The simulations used idealized models of galaxies
at $z=0.6$ and $I_{AB}=22$ (0.5 magnitudes brighter than our limit:
30\% of the CFRS sample is fainter than this)
with signal-to-noise ratios typical of the data used in this paper.
Galaxy models did not include bars, dust lanes, spiral structure,
or other irregularities.
The point-spread-functions (PSFs) are empirical and
derived from the data. They were assumed to be known exactly.
The seeing in the simulations was set at 0.67 arcseconds (FWHM),
roughly equal to the median of the real data.
Errors in centroiding, sky subtraction and PSF shifting 
were included.
The pixel scale for all simulations 
 was 0.20 arcseconds/pixel to match our
prime focus imaging.
 Three models were fit: a) disk only, b) bulge only and c) 
bulge-plus-disk model, and
the model with the smallest value of $\chi^2$
was accepted.

\subsection{ Measuring disk scale lengths}

Most of
the galaxies are expected to be disk-dominated (an expectation 
supported by the outcome of this study). Thus, a typical and 
important case is represented by a galaxy with
a small bulge fraction (B/T=0.20: approximately Sb) at $z=0.6$ and
$I_{AB}=22$ ($M_B=-21$). The bulge size was fixed according to
local relations (Sandage and Perelmuter 1990) at $R_{eff}=
0.28$ arcseconds (2.2 kpc). Galaxies similar to these are expected
to be common in the present sample.

The sky was measured and fixed prior to the fitting process and 
galaxy models were normalized to the number of counts in the
data being fit. The normalization was done after convolution
and within an aperture corresponding to optimum signal-to-noise
ratio (computed individually for each galaxy). 
Derived parameters were compared
 for the two cases where the sky was known
exactly and where the dispersion in the sky determination 
was 0.1\% (typical of the sky level uncertainty in the real data). 
The mean recovered parameters and the dispersion in those
parameters were not measurably different in the two cases. On the other
hand, errors of 1.0\% in the sky determination had a severe
effect on the determined parameters (and their errors) showing that 
the sky must be well-determined.

Figure 1 shows the results
of varying the disk scale length, $h$, (and thus the surface brightness
since the total luminosity is fixed) 
 from 0.10 arcseconds (0.8 kpc at $z=0.6$ or 0.5 pixels) 
to 1.6 arcseconds (12 kpc). 
  The fitted parameters were
disk scale length, axial ratio (the true value
was $b/a=0.75$), and position angle as well
as bulge fraction ($B/T$), bulge effective radius, and bulge axial
ratio (the true value was $b/a=0.9$). The position angles of 
the bulge and disk were assumed to
be coincident. 
The dispersion in the
recovered scale lengths is about 20\% and systematic errors are
small. The worst errors are at the largest scale length
which has the lowest signal-to-noise ratio per
unit area. The dispersion over the most relevant range in size 
($0.2< h < 0.4$ arcseconds) is 5-10\%.
showing that  meaningful measures of disk size can be made from
ground-based imaging for realistic-sized late-type galaxies at high redshift. 

 The typical dispersion in the recovered integrated disk
brightness is 15\% over the range $0.2< h < 0.4$ arcseconds
and this represents a reasonable uncertainty for late-type galaxies
in the sample of real galaxies.

 The dispersion in recovered bulge fractions ($B/T$) depends
on scale length but $\sim 95\%$ of the galaxies 
(with true $B/T = 0.2$) have a measured $B/T < 0.5$.
Over the range of the most relevant scale lengths ($0.2< h < 0.4$ arcseconds)
$\sigma_{B/T}=0.2$ is a good estimate of
the error. 

\subsection{ Measuring the fractional bulge luminosity}

A set 
of simulations was constructed of a galaxy at
$z=0.6$ with varying B/T (results in Figure 2). As $B/T$ varies, the bulge 
and disk sizes were also varied to agree with locally-defined surface brightness
constraints. The axial ratios of the bulge and disk were 0.9 and 0.75
respectively.
Over the full range of $B/T$, the worst dispersion
is $\sigma=\pm0.25$ in $B/T$ and the median dispersion is 0.14.
Distinctions between early and late-type systems can be made
but the further subdivision of late-type systems is difficult.
For galaxies like these we can construct perhaps three
classes: 1) early ($B/T> 0.75$), 2) mid-type ($B/T\sim 0.5$)
3) late-type ($0.0 < B/T < 0.3$). The corresponding
Hubble types are roughly 1) E, 2) S0-Sa, 3) Sb-Sd.

 The preceding experiment also indicates that is possible to
obtain useful measurements ($\pm 20\%$ or better)
of disk scale lengths for galaxies with $B/T < 0.6$. For earlier-type
galaxies the disk contributes much less light
and its size is very difficult to measure. 
Bulge
sizes can be measured reliably for galaxies only if $B/T \ge 0.75$. 

\subsection{Recovering the type distribution }

 It has been shown that model parameters can be recovered 
with reasonable accuracy from simulated datasets. In reality,
the galaxy type is not known beforehand and it must be demonstrated that
the fitting process does not result in a biassed estimate of
the population properties. This could happen, for example, if
a large fraction of mid-type galaxies were systematically
mis-classified as ellipticals. Only a crude classification is
attempted here ($B/T$ larger or smaller than 0.5) to determine
the fraction of disk galaxies that might have been misclassified.
Such misclassification might result in a biassed view 
of the properties of the disk galaxy population.

 In order to derive the completeness fraction for disks as a whole,
it is necessary to know the intrinsic distribution
of types and use this knowledge as input to the simulation
process. This information is not available but can be estimated
either from the local galaxy population or from the results of the
fitting of the high-redshift population. In Figure 3 a
simulated population  is shown (type is determined by bulge
fraction $B/T$). This generic distribution (a bias toward later
types) is similar to the magnitude-limited 
distributions shown
by Buta et al. (1994). The galaxies that make up this simulated
population cover a range in sizes and inclinations 
consistent with the real objects at high redshift and are
all simulated at $I_{AB}=22$.

The top panel of figure 3 shows the distribution of types used
as input to the simulation. 
 The second panel of figure 3 shows the recovered distribution
of $B/T$ after fitting the simulated population. The number of
galaxies in the extreme bins ($B/T\sim0$ and $B/T\sim1$) is large and has been
shared with the adjacent bins to improve the plots. The fraction of
galaxies that are correctly classified as early ($B/T<0.5$) or
late ($B/T>0.5$) was computed and found to be in excellent agreement
with the fraction expected if the error in determining $B/T$ was
$\sigma_{B/T}=0.2$. If the distribution
of types were uniform (all types equally likely) then 84\% of the
population would be correctly classified according to our
criterion. In the example population shown here the completeness
is higher: 88\% of the late-type galaxies are correctly classified.

The final panel
shows the measured distribution of $B/T$ in the real galaxy sample.
The number of galaxies with $0.0 <B/T< 0.1$ is 67 and these have again
been redistributed over 3 adjacent bins for plotting purposes. Evidently
the sample of real galaxies is dominated by late-type (disk-dominated)
systems. 
A distribution of this shape (heavily weighted toward late types) 
will result in a higher
fraction of correctly classified galaxies than would either a uniform
distribution or the distribution used as input to the simulation
shown in figure 3.
It could, therefore, be argued that 90\% or more of the
real galaxies are correctly classified. However, this is expected to
be true of galaxies like those in the simulations (smooth profiles
with no bars, dust lanes or any other structure). Any 
structural irregularities could modify this fraction
substantially. A comparison of the ground-based results with
those from $HST$ imaging (CFRS IX) demonstrates that such
abnormal structure does not severely affect the
present results. 

\subsection{Comparison with  results of HST imaging}

 There are 22 objects in common between the present sample
and the {\it HST} imaging work in CFRS IX . (A number
of galaxies at $z<0.5$ that did not appear in that work
are included here.)
Figure 4
shows a direct comparison of the parameters derived from ground-based
and {\it HST} imaging for the galaxies in common. 
Disk scale lengths are measured 
consistently from both datasets. The scatter in $B/T$ is large
but note that $\sim 67\%$ of the $B/T$ values lie within
$\pm 0.2$ of one another. 
Solid symbols are galaxies that were
classed as peculiar/asymmetric based on the  {\it HST} images. Several 
of the most serious differences between the measurements of $B/T$
from the two data sets are for these asymmetric objects. Perhaps
surprisingly, (at least for objects that are measured as
bulge-dominated from $HST$
and disk-dominated from the ground) 
the measured disk scale
lengths agree well for these  asymmetric objects.

 An estimate can be made of the effect that asymmetric objects
will have on the measurements in the present study. According to
CFRS IX, the  asymmetric objects constitute 30\% of the
high-redshift sample.  According to figure 4 about half of the
asymmetric objects have their fractional
bulge luminosities underestimated from the ground-based
data. (But bear in mind that the parameter $B/T$ may be poorly-defined
for asymmetric objects.) It seems
likely that objects of this type would tend to produce spurious
measurements that would be interpreted as high 
surface brightness disks and this is a possible source
of bias in measured disk properties.

 An examination of the scatter in the comparison of $HST$ and 
ground-based measurements suggests that the number of
objects with $B/T> 0.5$ as measured from the ground should be
increased by a factor of 1.6 to account for those
scattered to smaller $B/T$. Furthermore, 3/4 of the 
scattered objects are likely to be asymmetric/peculiar
(a propertiy that increases the likelihood that they
will be scattered to $B/T < 0.5$) and, therefore, likely
to be BNGs. This correction factor will be used in
/S4.4.

 In addition, the relative fractions of early and late-type
galaxies will be affected  by 
 the errors in ground-based measurements of $B/T$ and this will
scatter galaxies with $B/T > 0.5$ into the disk-dominated region,
with the result that
these will be classified and measured as disk-dominated
systems. The main effect will be to introduce a population of
small ``disks'' with high central surface brightness. The
effect of such a possible bias is not too severe simply because
disk-dominated galaxies are in the majority.

 The conclusion from figure 4 is that both disk sizes and bulge fractions are
measured well from both ground-based and space-based images {\it if the
galaxies are regular}. There is considerable dispersion in the estimates
of $B/T$ and a small number of serious discrepancies. 
The latter may be responsible for quantitative differences
in, for example, the amount of disk brightening as measured from {\it HST}
compared to CFHT and this issue will be addressed in \S 4.4.

\subsection{Summary}

 A number of conclusions can be drawn from the simulations and the
comparison of ground-based and space-based imaging.

 1) Disk scale lengths of galaxies of types 
$Sa-Sb$ and later with realistic sizes can be measured with reasonable
accuracy ($\pm 20\%$) and with no significant bias
using ground-based imaging with seeing of 0.7 arcseconds
(FWHM). Measurements
of disk sizes in earlier type galaxies (S0) are not reliable.

2) Estimates of bulge fractions ($B/T$) of typical-size
galaxies are possible. The median
dispersion in the simulations is $\sigma_{B/T}=0.14$ and the worst
(at $B/T=0.5$) is $\sigma_{B/T}=0.25$. The range $0<B/T<1$ can
be broken into perhaps 3 classes with this
level of discrimination.

3) The effective radii of spheroids can be measured reliably only
in early-type systems. For elliptical galaxies, the sizes can
be measured with an accuracy of $\sim 15\%$ over the physically
interesting range. 

4) Simulations of a complete population show that it is unlikely
that more than 10\% of our disk-dominated systems are misclassified
as bulge-dominated and it is the smaller disks that are most likely
to be classed this way.

5) A comparison with results derived from $HST$ imaging  shows that
the major source of serious error is due to un-resolved irregular
or asymmetric structure that cannot be detected  from the
ground. 

 Taken together these results indicate that it is possible to 
make useful quantitative measures of normal high redshift galaxies from
high-quality ground-based data. In particular, measures
of disk sizes can be made reliably for the majority 
of normal galaxies assuming the morphological mix is similar
to that observed locally. On the other hand, the effect of
non-regular structure must be taken into account.
It is
important to reiterate that these results are for high-redshift
galaxies near the magnitude limit of the CFRS survey.  It is
substantially easier to measure parameters for brighter galaxies
or those at lower redshift.

\section{RESULTS}

\subsection{ CFHT Data}

 Observations were obtained at
Canada-France-Hawaii telescope (CFHT) 1992 June 2-3 and
1994 May 7-8. During the
first run the prime focus CCD camera was used with pixel scale of 0.207
arcseconds per pixel and the seeing ranged from FWHM 0.6 to 0.8 arcseconds.
 A complete description  of the 1992 observations is
contained in CFRS I.
The field was observed in $V$ and $I$ filters for equal integration times
in a mosaic pattern which resulted in a varation of exposures
for a given galaxy from 2700 to 10800 seconds in each filter. The median
seeing was 0.68 arcseconds (FWHM) in the $I$ images and 0.64
arcseconds in the $V$ images.
 
 The 1992 observations were supplemented during a 2 night observing run
in May 1994. The stabilized imaging
spectrograph (SIS) was employed in its imaging mode with a pixel scale
of 0.0875 arcseconds per pixel. The tip-tilt active mirror was used with integration
times of 20ms on nearby $V\sim15$ guide stars  to partially correct
for fast image motion and improve the image quality (Le F\`evre et al. 1994).
Typically 60 minutes
of integration were obtained in the $I$ filter 
for 149 of the 195 candidate objects in
this field. Seeing ranged from 0.50  to
0.94 arcseconds (FWHM) with a median of 0.70 arcseconds. 
The field size was 3.0 arcminutes on a side.

 The median of the total integration times per galaxy from both runs was 4
hours. 
 The images were bias subtracted and flat-fielded with dome
flats. Finally, for each frame a sky flat
was created from a median of the program frames excluding the
frame to be flattened. The exclusion was done to remove the effect 
of each program galaxy on the flat that was used to process
its own image. Local sky values were measured for each galaxy and the mean
for each group of 10 nearby galaxies was adopted as the sky for all of
the 10 members of that group. The dispersion in sky for a group
on the  prime focus  imaging
was typically 0.1\% of the sky value for the prime focus images. Fewer
galaxies were available on each SIS image. Consequently, the error
in the sky was typically 0.3\%.

  The prime focus imaging has a wide field ($7^\prime \times 7^\prime$) and
typically 6-8 reasonably bright stars are included in the PSF,
 but the higher resolution (SIS) data covered
a much smaller field of view per CCD frame
and a correspondingly smaller number (1-3) of possible PSF stars
were available. As many as 30 images of a single galaxy were
available (in total from both instruments in both colours) 
and each image was extracted and retained as a separate datum, each with
a corresponding point-spread function to minimize the
effects of sampling and uncertainty in the PSF. 
The fitting of each galaxy was done
simultaneously over all of the individual images of that galaxy.
One set of parameters was fit and the ideal model of the galaxy was
then convolved with the PSF for each corresponding image.

\subsection{ The sample}

 Fitting was done on all 195 objects from
the CFRS 1415+52 field with secure spectroscopic identifications
(confidence class $\ge 2$, CFRS II).
 Of these, 49 were classified as stars on the basis of morphology
alone and the spectroscopic and morphological classifications agreed
in all cases.
Two of the three
AGNs in the sample were classified as stellar and the
third (lowest redshift) was flagged as a peculiar object 
on the basis of high surface brightness and extension.

 Stars and AGN were removed 
 leaving 143 galaxies
 with secure redshifts. Of these, 33 are
 early-type (bulge fraction $B/T > 0.5$) and 110 late-type
 ($B/T<0.5$) galaxies. Unacceptable fits resulted for 3 late-type systems
 (failure of disk parameters to converge to physically
 reasonable values).
 
 From the absence of any unresolved galaxies among the
objects that are certain or likely to be galaxies ,
it is estimated that the frequency of galaxies with point-source
morphology is less than 2.5\% at the 99\% confidence level.
Crampton et al. (CFRS V) found that 7 of 591 galaxies (1.2\%)
were indistinguishable from point sources as judged by  
a compactness parameter.

 The galaxies without secure spectroscopic identifications 
were also fitted and 
 have morphologies and colors typical
of the other galaxies in the sample. 
  
\subsection{Distributions of inclination and color}

 The distributions of axial ratios (shown in figure 5) for
 disk-dominated ($B/T < 0.5$) and bulge-dominated ($B/T > 0.5$)
 galaxies are significantly different. This is expected if the intrinsic
 flattenings of the two populations differ and the result
 suggests that we are making meaningful classifications
 based on $B/T$. The flat distribution of axial ratios also
indicates that a bias in central surface brightness is unlikely.

 The distributions of rest-frame color,  $(U-V)_{AB,\circ}$
(obtained by interpolatation among the spectral
energy distributions of Coleman, Wu, and Weedman 1980, see CFRS VI
for details)
are shown in Figure 6. Galaxies with $B/T > 0.5$ have a redder
distribution of rest-frame colors than those galaxies whose light profiles
are dominated by a disk. This is further support for the
$B/T$ classifications although 
there is evidently a small 
population of apparently bulge-dominated galaxies with very blue colors
(these are the ``blue-nucleated galaxies'' identified in CFRS IX, see below). 

\subsection{Color versus morphology }

Figure 7 shows the
relation between $B/T$ and $(U-V)_{AB,\circ}$
at low (upper panel) and high (lower panel) redshifts. 
Dotted lines at $(U-V)_{AB,\circ}=1.4$
(color of present-day $Sbc$ galaxy) and $B/T=0.5$ divide the sample
into early and late-type galaxies. The sample is dominated by late-type
galaxies with blue colors ($(U-V)_{AB,\circ}<1.4$, $B/T < 0.5$) 
and has a small
number of normal early-type systems (redder and with $B/T > 0.5$). 
In addition to these expected subsets of the galaxy population, there are
a total of 16 objects in the region $(U-V)_{AB,\circ}<1.4$, $B/T > 0.5$
(3 at low redshift and 13 at high redshift).
 These objects would need to have bulges substantially
bluer than those of nearby galaxies in order to occupy this region
of the color-morphology plane. We identify these
galaxies with the ``blue-nucleated galaxies'' (BNGs)
discussed in CFRS IX.
The apparent frequency of the BNG phenomenon (which was shown to
be associated with peculiar/asymmetric structure in CFRS IX) 
is $6^{+6}_{-3}\%$ of the sample at $0.2 < z < 0.5$ and
 $14 \pm 4\%$ at $0.5 < z < 1.2$ (but see below). The change in frequency 
with redshift hints
at evolution, although disentangling luminosity and redshift effects
is difficult in this magnitude-limited sample.

 It is important to estimate the effect of asymmetric structure
on these results. A crude correction factor was derived in \S 3.4
to account for the scattering of galaxies with $B/T > 0.5$
into the region $B/T < 0.5$. The correction was derived from
very few galaxies and is thus uncertain.  
The fractions of 
galaxies with $B/T > 0.5$ in the CFHT and $HST$ samples are
0.23 (33/143) and 0.41 (13/32) respectively, a significant difference. 
Applying the correction factor of 1.6 indicates that $\sim 20$ galaxies
with $B/T > 0.5$ may have been scattered  into the $B/T < 0.5$. Then
the corrected fraction for the ground-based sample is 0.37 (53/143),
comparable to the $HST$ sample. Of these 20, 75\% are
likely to be asymmetric and we will assume that these are all BNGs. Then
the corrected freqeuncy of BNG's would be 26\% at high redshift
(compared to 30\% from $HST$)
and 13\% at low redshift, roughly a doubling in the rates of
occurrence due to this correction. This result is plausible
and brings the ground-based and $HST$ results into close
agreement.
 This analysis demonstrates that no significant
contradiction exists between the ground and space-based results
and that systematic errors due to unresolved structures are likely
to be responsible for any apparent differences.

\subsection{Disk scale lengths}

 Figure 8 shows the relation between scale length and disk
luminosity. 
 The relation for galaxies with $z< 0.5$ (median $z=0.28$) 
does not differ substantially
from the Freeman (1970) constant surface brightness relation (dotted line)
or the samples of Kent (1985) and van der Kruit (1987) (small symbols).
 The mean inclination corrected central surface brightness in
the rest frame 
is $\mu_{AB}(B)=21.3 \pm 0.25$ (standard error of the mean).
This agrees with the findings of Colless et al. (1994) that little
evolution in disk properties has occurred at these redshifts.
 The same relation for 70
galaxies with $z > 0.5$ (median $z=0.73$) is shown on the
lower panel of Figure 8. Nearly all of the galaxies lie away
from the fiducial realtion. The mean  rest-frame central
surface brightness is $\mu_{AB}(B)=19.8 \pm 0.1$, approximately
1.6 magnitudes brighter than the Freeman value.

  A comparison of the disk brightening with the value derived
from $HST$ imaging (CFRS IX) shows good agreement. The $HST$
value ($\mu_{AB}(B)=20.2 \pm 0.25$ magnitudes) differs at the
1.5 $\sigma$ level but this discrepancy is easily explained
as a result of small systematic errors  due to scatter in the $B/T$
measurements due to asymmetric structure (see \S3.4).

\subsection{The disk-size function}

 It is interesting to examine the space density of disks as a function
of size and compare this ``size function'' at high and low 
redshift. In principle, the present sample is magnitude limited
with no selection related to galaxy size (CFRS I). 
 Weighting by the inverse of the accessible volume
(calculated as in  CFRS VI) 
allows us to derive the space density
as a function of scale length within the apparent magnitude limits
of the survey. 

 The space density of disks as a function of scale length $h$
(the size function) is shown in Figure 9. At high redshift 
the number of small disks (scale lengths $1< h < 4$ kpc) with 
luminosities greater than $M_{AB}(B)=-20$ 
is  $2.6 \pm 0.2 \times 10^{-3}$ Mpc$^{-3}$ compared to
$0.3\pm 0.2 \times 10^{-3}$ Mpc$^{-3}$ at low redshift,
a substantial difference.
The number of disks at $0.2 < z < 0.5$ 
in this same size range but more luminous than
$M_{AB}(B)=-18.5$ is $2.9 \pm 0.8 \times 10^{-3}$ Mpc$^{-3}$. Thus
it is necessary to reach only $\sim 1.5$ magnitudes
lower in luminosity at low redshift to find a space density
of small disks directly comparable to that at high redshift.
This simply demonstrates that a scenario of
moderate luminosity evolution of individual disk galaxies reproduces
both the observed evolution of the luminosity-size relation
and the observed change in the space density of small luminous
disks between $z \sim 0.3$ and $z \sim 0.75$. No change in numbers
of galaxies is required. This is directly analagous to the situation
with the luminosity function of blue galaxies (CFRS VI) where it is necessary to
reach down a magnitude or so deeper into the luminosity function
at $z\sim 0.3$ to get the same space density as at $z\sim 0.7$.

\section{DISCUSSION}

\subsection{ Luminosity function---morphology relation}

 It was found (CFRS VI) that the 
evolution of the galaxy luminosity
function is dependent on galaxy color. The 
luminosity function of the reddest galaxies
(redder than present-day $Sbc$) shows very little
change, either in number density or
characteristic luminosity. The luminosity function of the blue galaxies
shows strong evolution. 

 The luminosity function provides a statistical description
of the galaxy population. As such, it contains no
direct information about the evolution of individual
galaxies. The observed luminosity function is consistent
with the fading of each of its constituent galaxies
by $\sim 1$ magnitude between $z=0.75$ and $z=0.3$. It is also
consistent with the existence of a new population of objects at $z=0.75$
with a luminosity distribution similar to the normal
population but with a space density a factor $\sim 3$ times
as high as the population that exists at $z=0.3$. More
complicated scenarios including color evolution  whereby
galaxies cross the red-blue divide can also be constructed
that are consistent with the luminosity function behavior.
In other words, the behavior of the luminosity function
can be described equally well by a number
of evolutionary schemes.
Morphology provides further
constraints on the evolution of individual galaxies.

 The relation between color and morphology indicates that
the color selection 
applied in the luminosity function derivation is similar to
our purely morphological cut at $B/T=0.5$, between disk-dominated
galaxies and bulge-dominated galaxies. In particular, disk-dominated
systems form the majority of our galaxies and are nearly all
bluer than the nominal red-blue division. Therefore, it is
reasonable to identify those galaxies with $B/T<0.5$ with
the blue population whose luminosity function is evolving
rapidly. (The BNG galaxies are a second component of
the blue population.)

 Galaxy disks at $z<0.5$ have a size-luminosity relation
that is consistent
with no evolution compared to the  
local relation. The luminosity function at these
same redshifts agrees reasonably well with that determined locally
(CFRS VI).
The morphological
result agrees basically with the luminosity function result.

 At $z>0.5$ the morphological analysis reveals 
an apparent brightening of a large fraction of disk galaxies
relative to the local population. This represents
a dramatic change in the properties of these galaxies.
 The evolution in the luminosity function shows
a brightening of the characteristic luminosity
by $\sim 1$ magnitude.
This rough correspondence of the morphological and
luminosity function analyses suggests 
 that one of the components of the evolution in the LF is due to
the brightening of individual (nearly normal) disk galaxies rather than to
a dramatic change in number density due to
merging or the disappearance of an exotic population. 

 These results, taken at face value, imply that
{\it nearly all} disks have substantially enhanced rates of star-formation
compared to present-day galaxies whereas local observations
 indicate
that the star-formation history of nearby disk galaxies is a strong function
of their present-day Hubble type (Kennicutt, Tamblyn, \& Congdon 1994).
The ratio of present to past-average star-formation rates varies from
$<$1/10 to unity for $Sa$ to $Sc$ galaxies respectively. This implies
that some fraction of disks should have higher surface brightness in our
high-redshift sample but not all. In fact, those with constant star-formation rates
are expected to have lower surface brightness at high redshift. 
Why do we see, almost exclusively, enhanced disk surface brightness?
Selection is probably part of the answer: if part of the population has higher
surface brightness (and therefore higher luminosity) 
then that part of the population will be 
preferentially observed in a magnitude limited sample. The
full answer is likely to be more complicated and
 a detailed reconciliation of these results needs to be done. 

\subsection{Blue nucleated galaxies}

 In addition to the observed disk-brightening, a second
evolutionary effect appears to be the emergence of
the ``blue nucleated galaxies'' (BNGs) identified
in CFRS IX. It has been argued (\S 4.4) that the frequency of
this phenomenon in the ground-based sample should be corrected
for the effects of abnormal structures. 
The derived correction factor leads to the estimate that
BNGs make up 26\% of the
sample at $z>0.5$ (compared to 30\% of the $HST$ sample),
and 13\% at low redshift. The difference in frequency is suggestive
but not very significant. 

 The BNG phenomenon has been shown to be strongly associated with
asymmetric structure and  mergers/interactions (CFRS IX) but
these perturbations are difficult or impossible to detect from
ground-based data so we cannot confirm this association.
 The Virgo cluster contains a population of spiral galaxies
with large, star-forming bulge-like central concentrations
(van den Bergh, Pierce, \& Tully  1990) that may be  
a related phenomenon. This class of objects makes up
a significant fraction of the total population and is probably 
a second major component of the evolution of the galaxy population
at high redshift.

\subsection{Comparison of our results with previous HST imaging}

 Griffiths et al. (1994a) give half-light radii for 38 disk
(or probable disk) galaxies brighter than
$I=22$ ($I_{AB}=22.5$). These
measures were converted to scale lengths (divided by 1.67) and
compared to our own measurements.  
At a given magnitude there is little difference in sizes between
the present results and those of Griffiths et al. 
 Their apparent excess of compact objects compared to
no-evolution models is qualitatively consistent with
our results.

 Classifications are presented by Griffiths
et al. (1994b) who conclude that the morphological mix
of normal faint galaxies to $I=22$ is similar to that observed locally,
but the number of anomalous or peculiar objects 
is larger than expected,
reaching 40\% at the faint limit. This point is
elaborated upon by Glazebrook, Ellis, and Santiago
(1995) who show that the counts of
normal galaxies are consistent with models of the local
population but that the slope of the number-magnitude
relation for irregular/peculiar galaxies is 
steeper. They propose that these anomalous objects constitute the
rapidly-evolving component.
 Forbes et  al (1994) found weak evidence that galaxies showing
signs of interaction have bluer nuclear colors than the general
population.

 Driver et al. (1995) show that the counts of ellipticals and
early-type spirals are consistent with no-evolution models but argue
that the strong excess in the counts of late type spirals and
irregulars may be produced by adopting a dwarf-rich local
luminosity function and assuming that much of the late-type
population experiences a dramatically elevated level of star-formation 
at $z\sim 0.5$.

 The present work is consistent with these previous results. Our disk
sizes are supported by the work of Griffiths et al (1994a). 
We find
a population of compact objects (large $B/T$) with blue colors (BNGs) although
we do not have sufficient resolution to detect mergers/interactions 
reliably.  The association of the BNG phenomenon with interactions
is, however, demonstrated in CFRS IX, thus confirming the Forbes et al. result.
 The work by Driver et al. (1995) has a number of strong similarities to
our own conclusions.

 The ground-based imaging lacks sufficient resolution for a meaningful
estimate of the fraction of  irregular/peculiar galaxies discussed
by Glazebrook, Ellis, and Santiago (1995) although $HST$
imaging a (CFRS  IX) indicates that 30\% of high-redshift ($z>0.5$) galaxies
have display structural asymmetry. That fraction counts as asymmetric/peculiar
those galaxies with more than $5\%$ of the total galaxy flux contained
in an asymmetric component. The peculiar fractions
agree if each is taken at face value, but  the lack of a consistent
definition of peculiar/asymmetric structure among different
groups makes it unclear whether frequencies of the same types of
structures are being compared.

 In addition to the concern that different groups may be
defining asymmetric, peculiar, or irregular structure differently,
there is a complete lack of analogous measurements in the local
galaxy population. Without that context, the frequency of
such structures in the faint galaxy population is difficult 
to interpret.

\subsection{Summary}

 The disks of late-type galaxies at high redshift ($0.5<z<1.0$)
have a higher mean surface brightness than either their local
counterparts (Freeman 1970) or the galaxies in the present
sample at $0.2<z<0.5$. 
 The majority of the disks that we detect are apparently forming stars at a
higher rate at high redshift than at low by a factor of 2-4.

 The correspondence of the change in disk central
surface brightness and the behavior of the luminosity
function of blue galaxies (CFRS VI)
suggests that moderate luminosity evolution of
normal disk galaxies
is a major evolutionary effect on the galaxy 
population. 

 At high redshift, 14\% of the galaxy population are classified
as ``blue-nucleated galaxies'' but the true frequency of occurrence
is probably larger by a factor of 2 when the effect of
systematic errors is included.
There is a hint (at the $1\sigma$ level)
that the frequency of this BNG phenomenon increases with redshift. 
 This phenomenon is probably an important second component 
in the evolution of the galaxy population.

\subsection{Outlook}

 The multi-variate galaxy luminosity function\\
 $\phi(L,z,colour)$ 
shows that strong evolution is occurring in the blue galaxy population
at high redshift (CFRS VI).
 The present results illustrate the progress that can be made when
additional morphological dimensions are added to the general multi-variate
galaxy distribution function\\
 (e.g., $\psi(L,z,h,(U-V)_{disk},r_{eff},
(U-V)_{bulge},B/T$))
of which the luminosity function is one projection.
A discussion of a second projection, 
 the ``size function'', gives additional weight to the
suggestion that luminosity evolution of individual disk galaxies is a major 
component of the evolution of the population. 

The evolution of
an individual galaxy can be described by a vector with
components $\Delta L_{disk},\Delta (U-V)_{disk},\Delta h, \Delta L_{bulge}$,
etc., and the sum of these individual vectors gives the
corresponding changes in the distribution function $\psi$.
Any theory of galaxy evolution gives specific predictions for
the evolutionary vectors and  thus the change in the distribution
function. {\em All of the predicted changes must  be observed
if the  evolutionary model is correct.}
 A sample with a well-determined selection
function is required to compute the distribution function (which has units 
of absolute density, i.e., Mpc$^{-3}$).  The present
ground-based sample begins to demonstrate the power of this
approach to morphological analysis but detailed measurements of
the morphological parameters of a sample 
much larger than that presented here, and with much higher-quality imaging
(only possible with {\it HST}) will produce more stringent constraints
on models of galaxy evolution than those that exist at the present time.

\newpage

\centerline{Figure captions}

Figure 1: Disk scale lengths recovered from simulations. A range
of disk sizes was used as input but the other parameters were
fixed: bulge $R_{eff}=0.28''$, fractional bulge luminosity
$B/T=0.2$, disk axial ratio =0.75, bulge axial ration 0.90.
Disk and bulge position angles were assumed to be identical
in both the simulations and the fits. Galaxies similar to this
are expected to be common in our sample. The signal-to-noise
ratio and scale sizes are typical for a galaxy at $z=0.6$
and $I_{AB}=22.0$ (0.5 magnitudes above our limit) and so represent
a difficult but important case. The scale lengths are recovered
well. The error bars on the points corresond the the average
error bar output by the fitting program and these errors are
confirmed to be reliable by comparison with the dispersion 
in recovered parameters.

Figure 2: Simulations of a typical high-redshift galaxy near our limit
(similar to that in figure 1). We simulate a range in fractional
bulge luminosity ($B/T$) from $B/T=0$ (pure disk) to $B/T=1.0$)
(elliptical galaxy). The bulge and disk sizes vary to comply
with locally determined relationships between size and luminosity.
 We recover the true values of $B/T$ reasonably well but we find
from simulations such as these that we can measure disk sizes
reliably only for galaxies with $B/T < 0.6$ and can measure
bulge sizes reliably only for those galaxies with $B/T > 0.75$.

Figure 3: The top panel shows an input simulated galaxy population,
the second panel shows the distribution of $B/T$ recovered from the
simulation, and
the final panel shows the distribution of $B/T$ measured for
the actual CFRS data. It is likely that more than 90\% of 
normal galaxies are correctly classified as early ($B/T< 0.5$)
or late ($B/T > 0.5$).

 Figure 4: A direct one-to-one comparison of ground-based measurements of
galaxies in the present sample with measures from HST WFPC2 imaging.
 In general, the correspondence is good. Asymmetric/peculiar galaxies
(filled symbols) are more likely to to be subject to serious 
measuring errors than normal galaxies (open symbols).

Figure 5: The distribution of fitted axial ratios for the real sample.
The two distributions (for bulge- and disk-dominated galaxies)
agree with expectations since the intrinsic distribution of
bulge axial ratios avoids values as flat as disks. 

Figure 6: Colour distributions in the rest frame for bulge- and
disk-dominated galaxies. Together with figure 3 this shows that
our purely morphological discrimination pick out two populations
with very different colors.

 Figure 7: The relation between color and measured bulge fraction.
 The normal galaxy sequences are disk-dominated ($U-V< 1.0$, $B/T \sim 0.2$),
elliptical/S0 ($U-V> 1.5$, $B/T > 0.6$), and  mid-types
($U-V\sim 1.5$, $B/T \sim 0.5$). The objects  with
($U-V< 1.4$, $B/T > 0.5$) are anomalous. They evidently have
compact components that are too blue to be normal bulges.
There objects are referred to as ``blue-nucleated galaxies''
(BNGs).

 Figure 8: The relation between disk size and luminosity for 
galaxies with $B/T < 0.5$.   At low
redshift the galaxies scatter about the Freeman (1970)
relation (dotted line). At high redshift a large fraction of
the galaxies are too bright at a given size. The mean
disk central surface brightness at high redshift is
sgnificantly higher than the Freeman value. Typical error
bars for the total disk luminosity are shown in the
lower left of the figures. Samples from
Kent (1985) and van der Kruit (1987) are plotted as small
squares and triangles respectively. 

 Figure 9: The space density of disks as a function of size in
bins of width 2 kpc. The bins are centered at 2,4,6, and 8 kpc
but arbitrary offsets are applied to make the plot more
legible.
 This figure shows that the space density of small ($1< h < 4$ kpc)
disks (which consitute the majority) is constant with redshift. There is
a surplus of luminous ($M_B<-20$) small disks at high redshift (with a median
luminosity of $-21.0$) relative to the same size range at low redshift.
This surplus is accounted for by the number density of lower
luminosity disks ($M_B>-20$; median luminosity $-19.7$) at lower
redshift ($0.2<z<0.5$). 

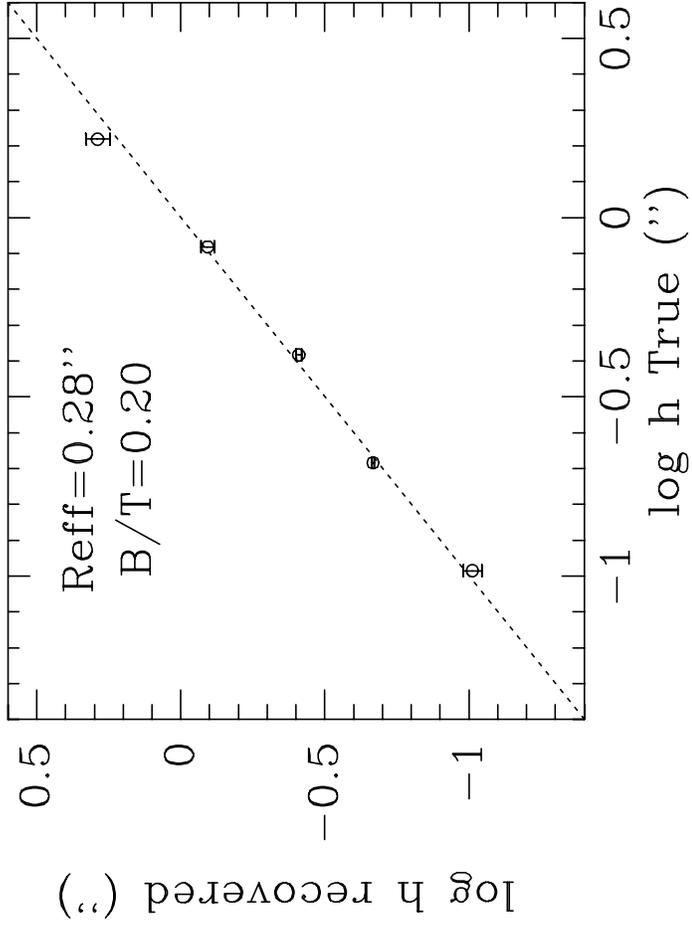
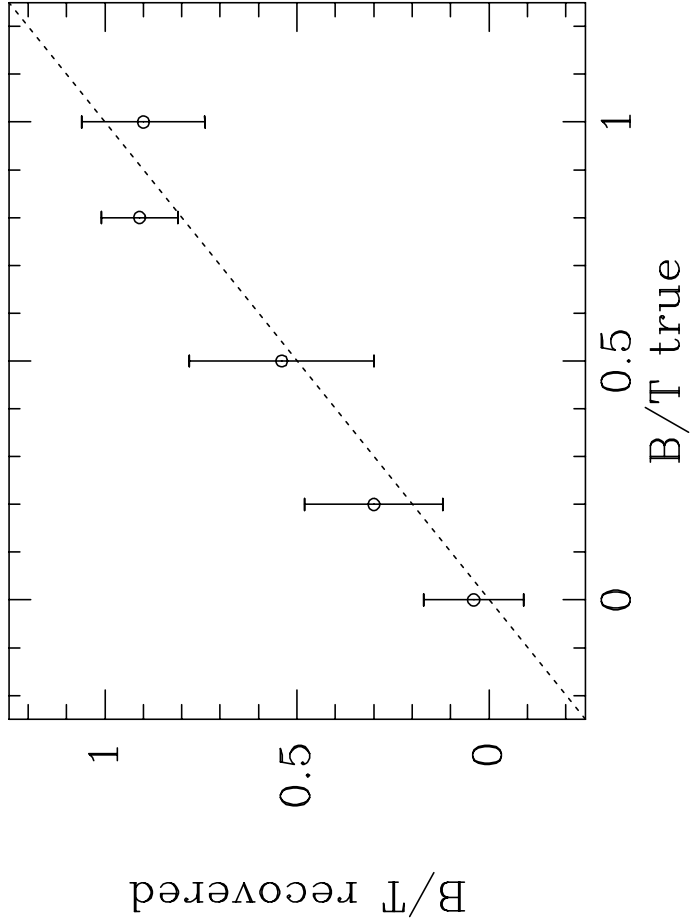
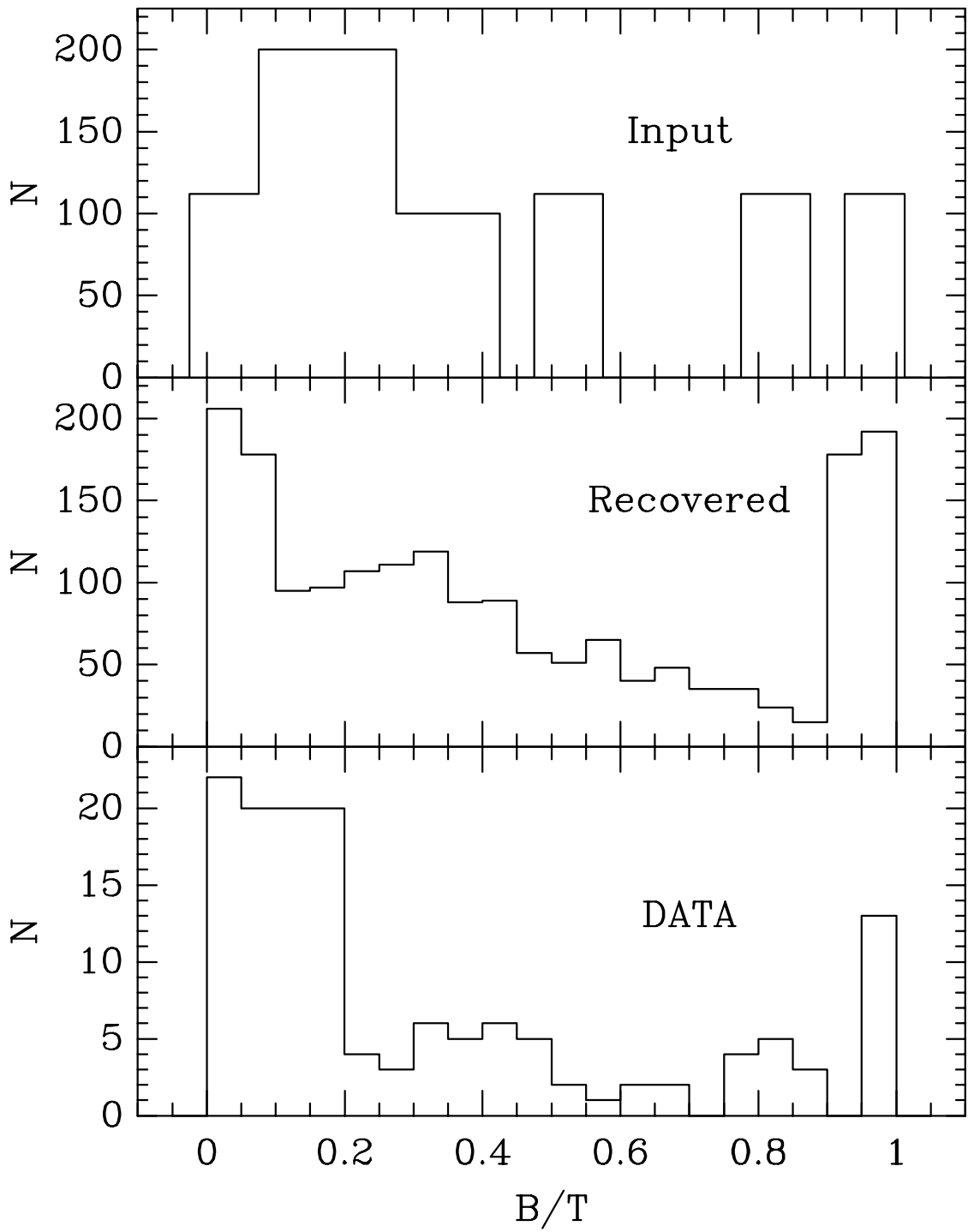
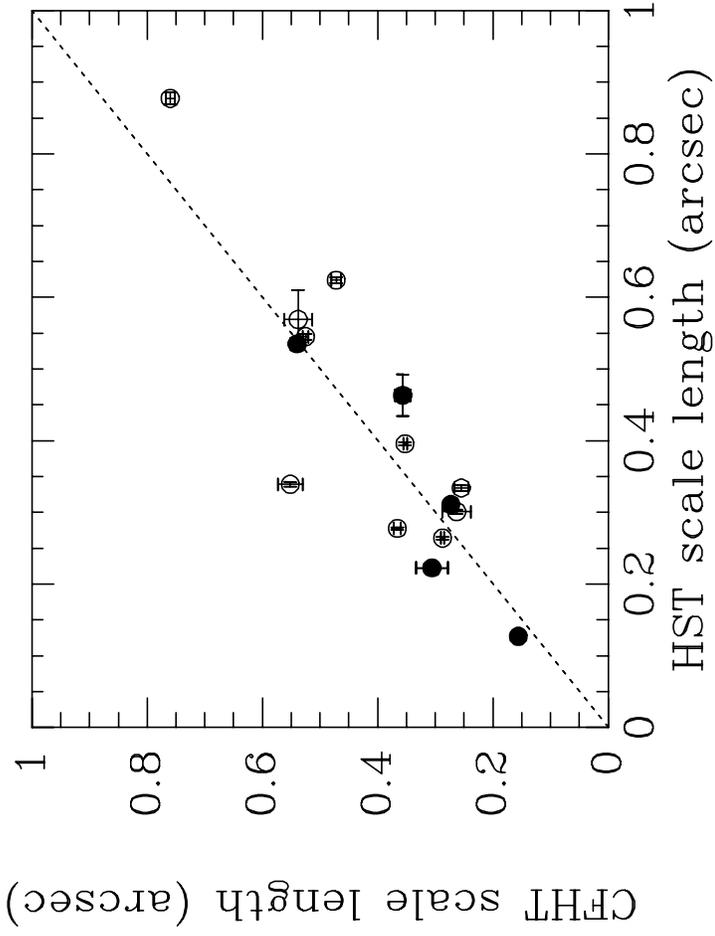
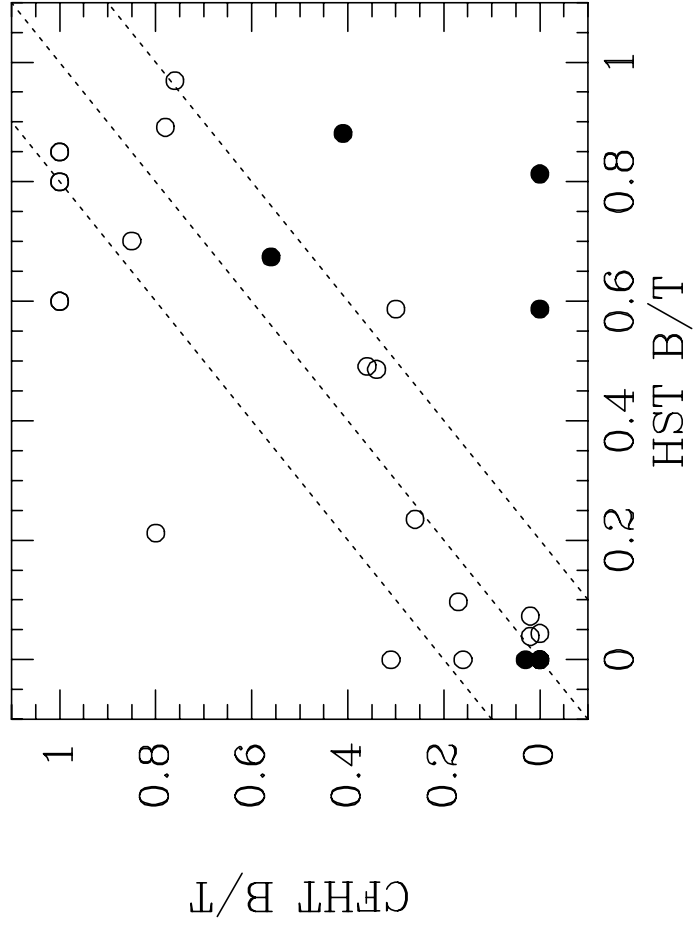
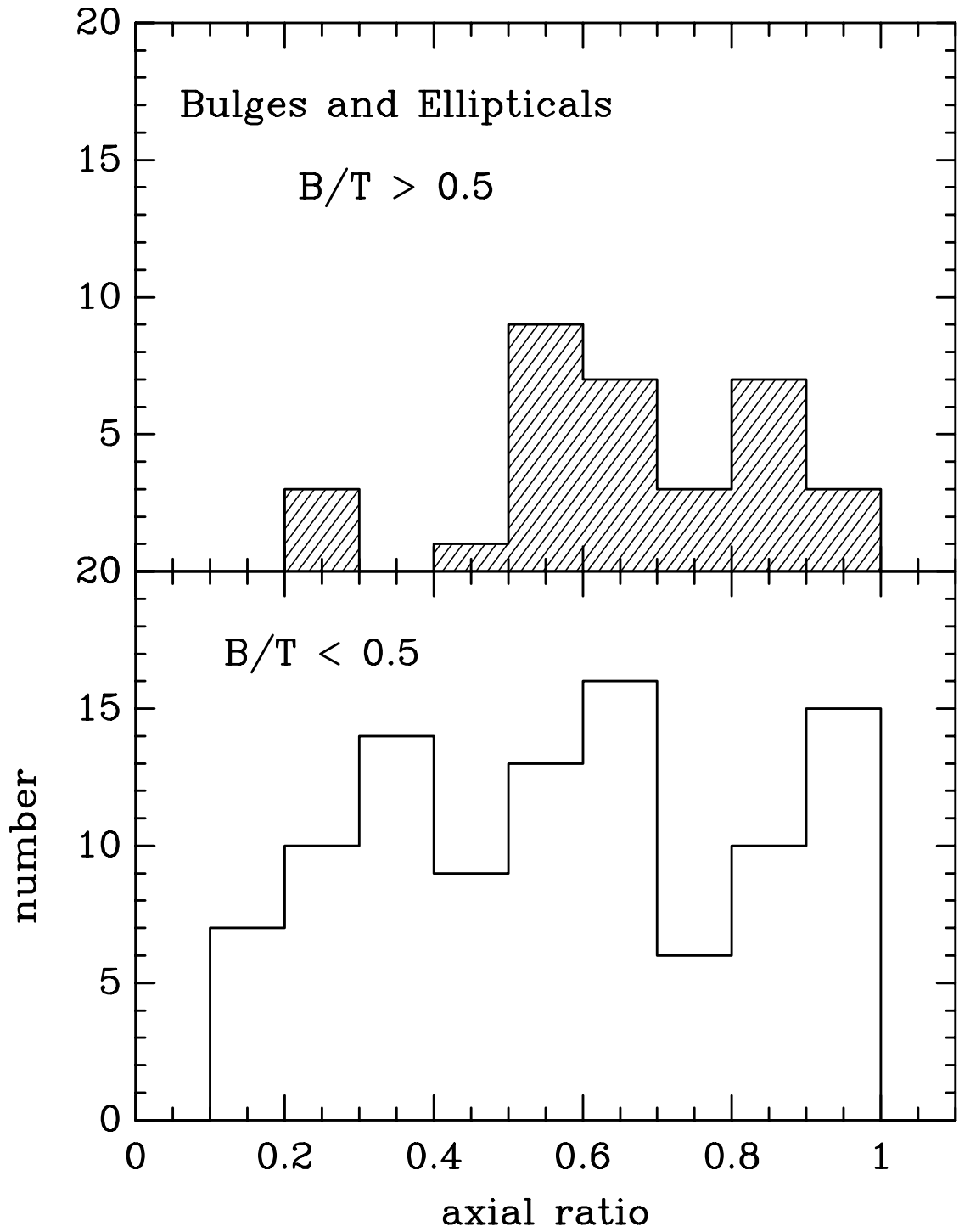
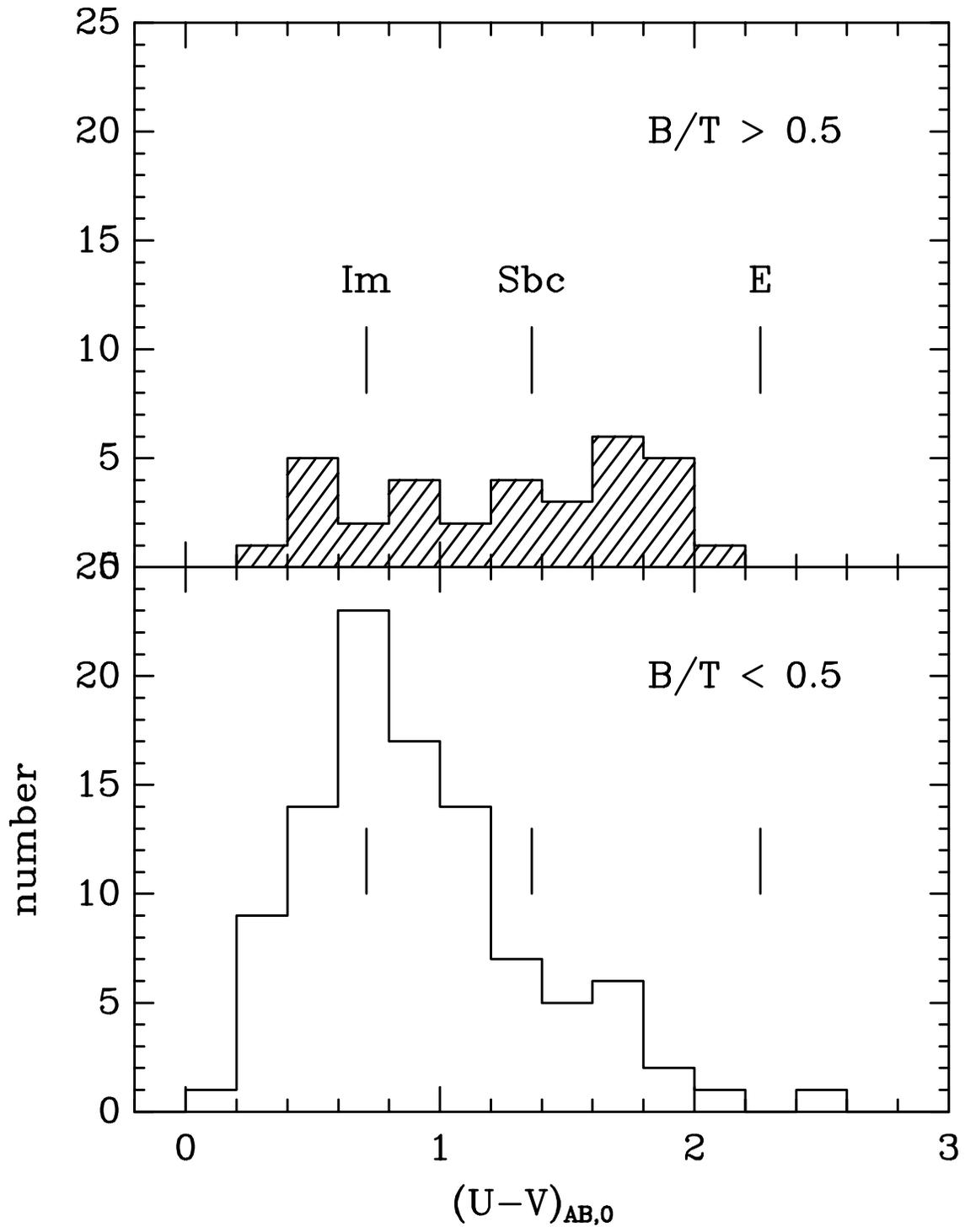
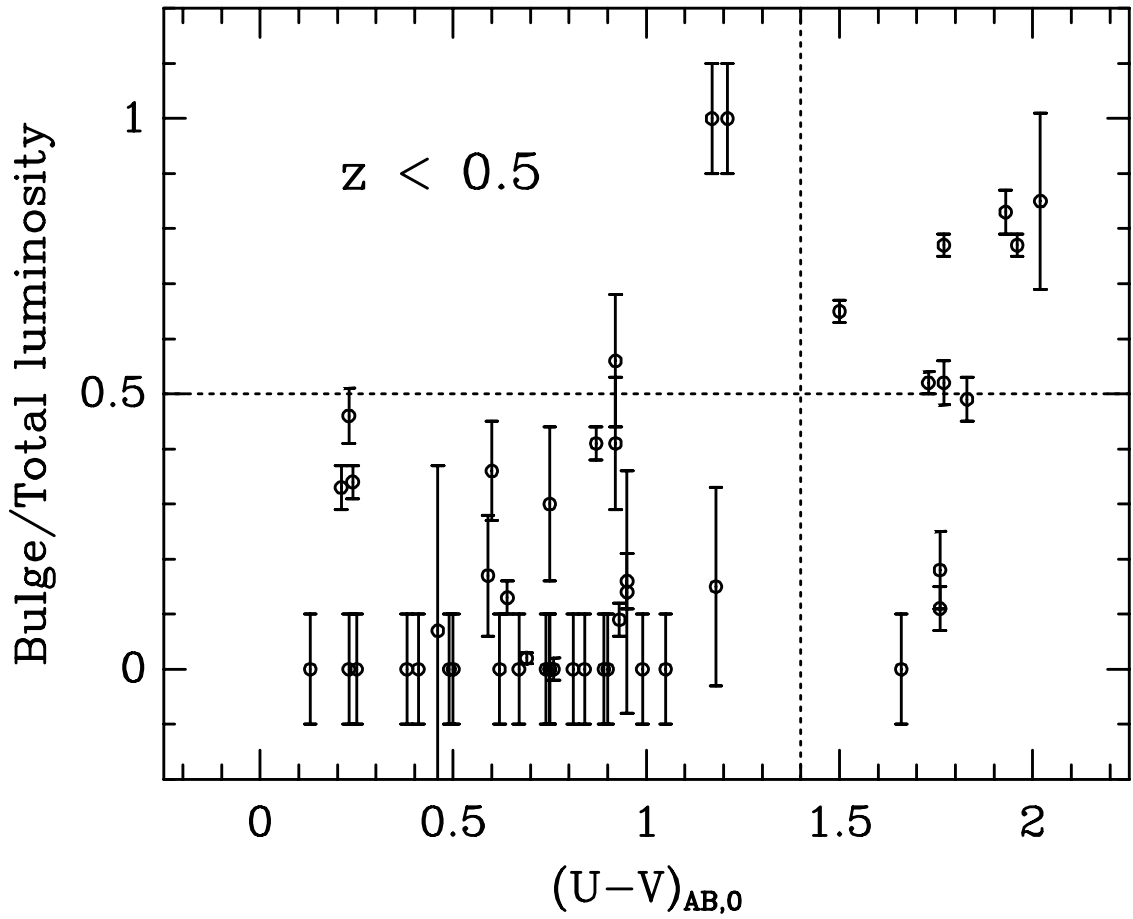
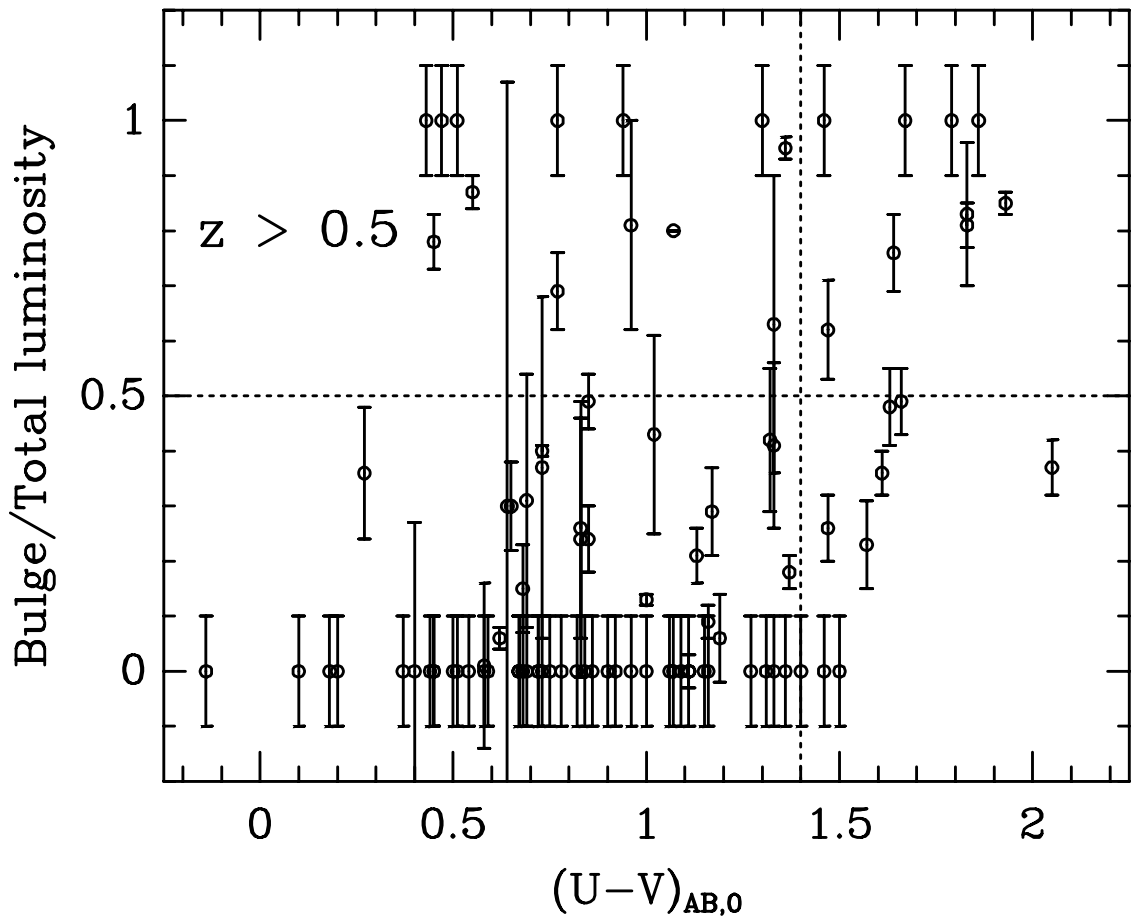
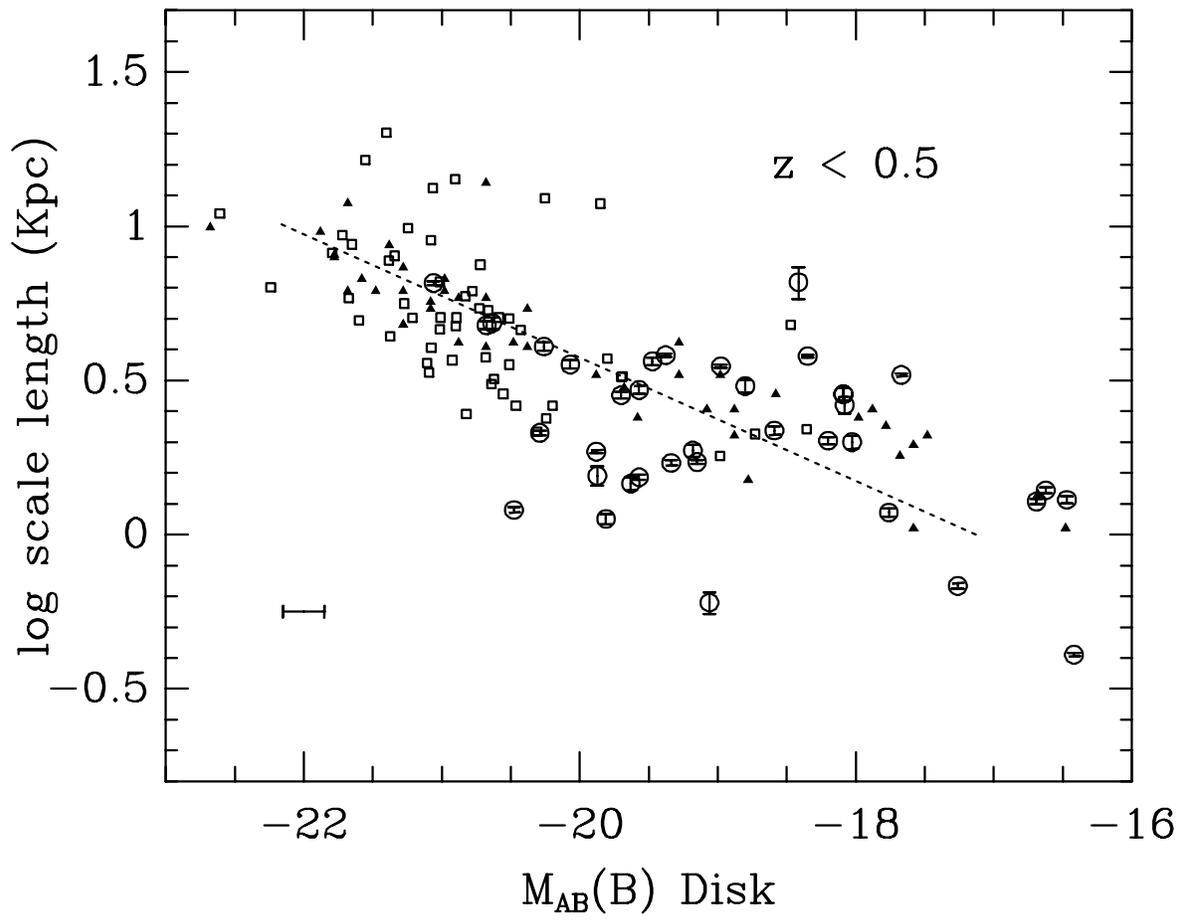
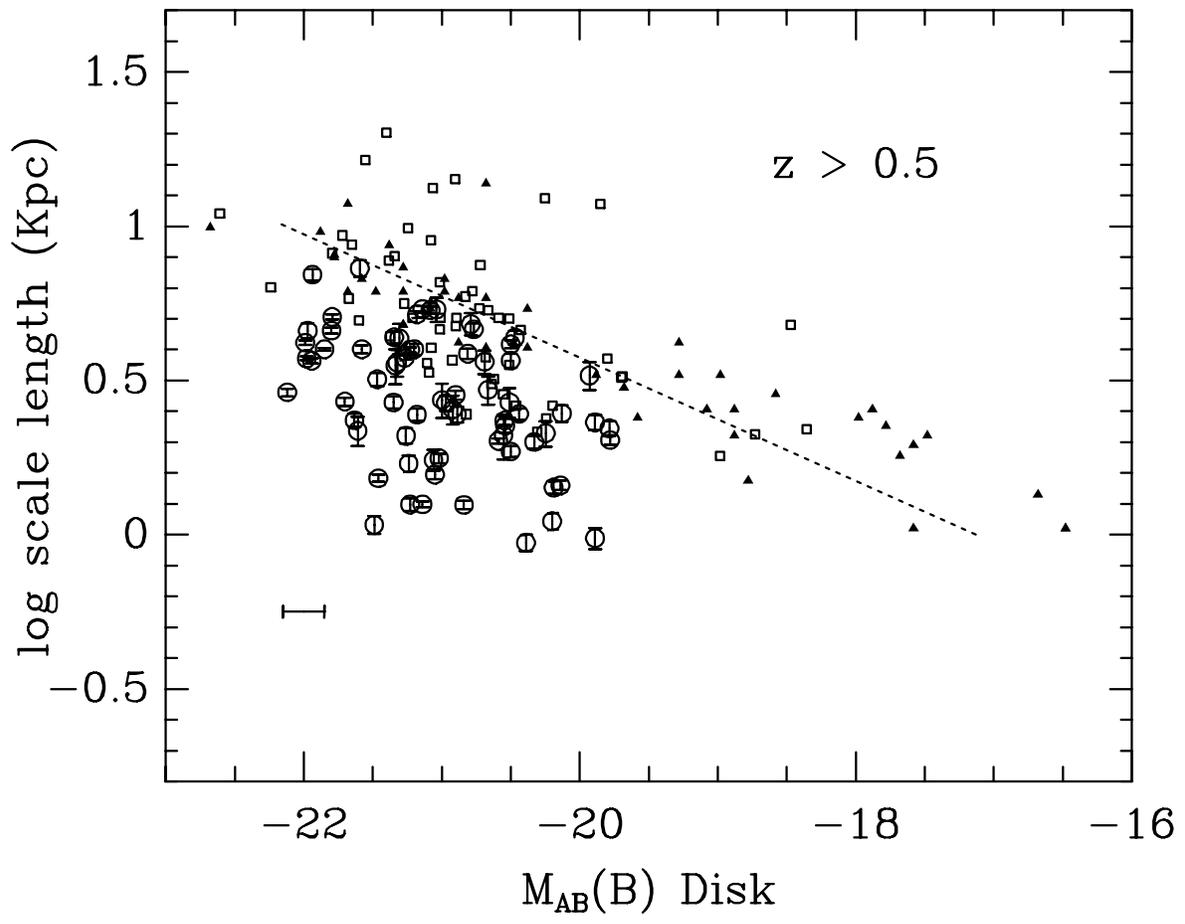
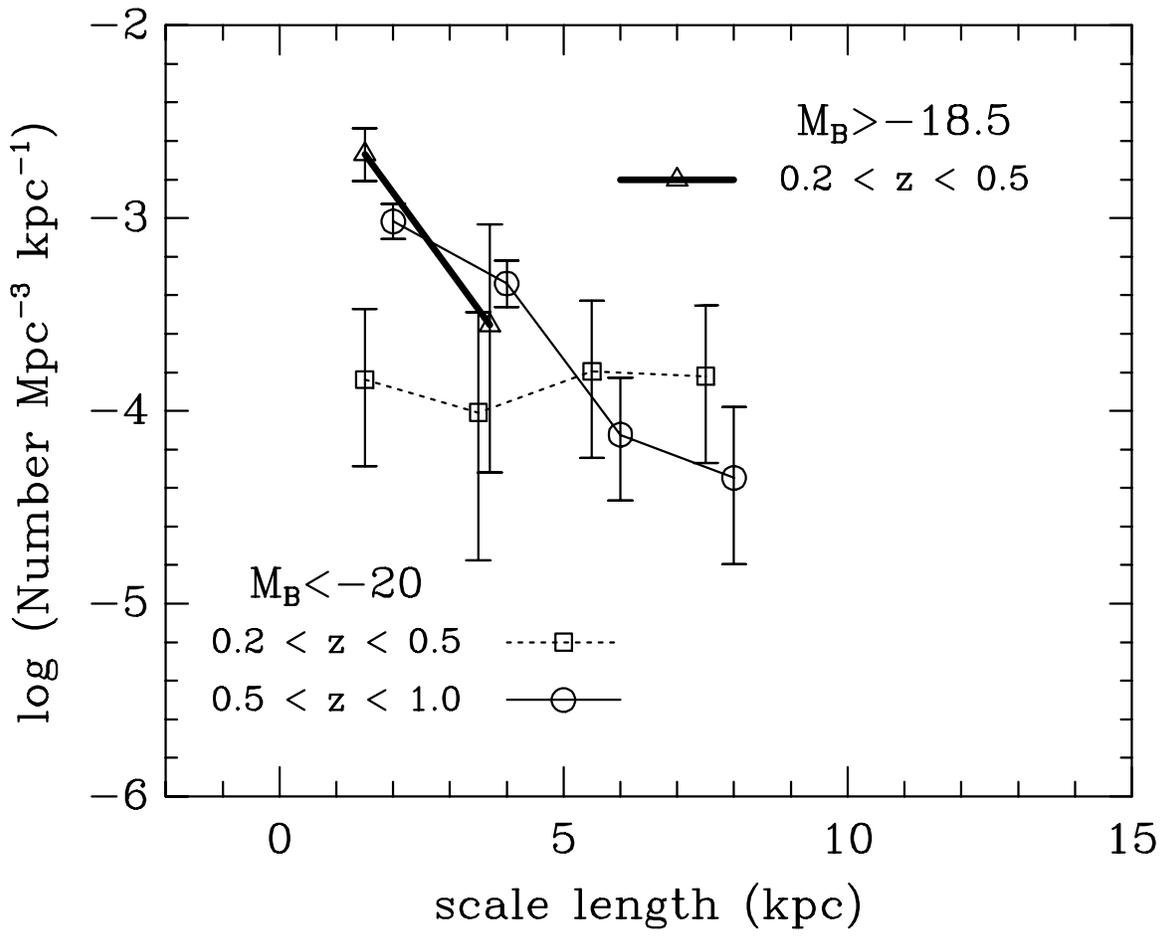
\end{document}